\documentclass{PoS}

\title{Studies of methanol maser rings}
\ShortTitle{Studies of methanol maser rings}
\author{\speaker{Anna Bartkiewicz}, Marian Szymczak\\
        Torun Centre for Astronomy, Nicolaus Copernicus University, Poland\\
        E-mail: \email{[annan;msz]@astro.uni.torun.pl}}
\author{Huib Jan van Langevelde\\
        Joint Institute for VLBI in Europe, Dwingeloo, The Netherlands\\
        Sterrewacht Leiden, The Netherlands\\
        E-mail: \email{langevelde@jive.nl}}
\author{James Michael De Buizer\\
        Stratospheric Observatory For Infrared Astronomy, USA\\
        E-mail: \email{jdebuizer@sofia.usra.edu}}
\author{Ylva Pihlstr\"om\\
        Department of Physics and Astronomy, University of New Mexico, USA\\
        National Radio Astronomy Observatory, Socorro, USA\\
        E-mail: \email{ylva@unm.edu}}

\author

\abstract{We present the results of studies of a new class of 6.7\,GHz 
methanol maser sources with a ring-like emission structure 
discovered recently with the EVN. We have used the VLA to search for 
water masers at 22\,GHz and radio continuum at 8.4\,GHz towards a sample of
high-mass star forming regions showing a ring-like distribution of
methanol maser spots.
 Using the Gemini telescopes we found mid-infrared (MIR) counterparts
of five methanol rings with a resolution of 0."15. The centres of methanol
maser rings are located within, typically, only 0."2 of the  MIR emission
peak, implying their physical relation with a central star. 
These results strongly support a scenario wherein the ring-like structures
appear at the very early stage of massive star
formation before either water-maser outflows or H\,{\small II} 
regions are seen.}

\FullConference{10th European VLBI Network Symposium and EVN Users Meeting:
VLBI and the new generation of radio arrays \\

                 September 20-24, 2010\\

                 Manchester, UK}

\begin{document}

\section{Introduction}
Methanol maser emission at 6.7\,GHz is a well known tracer of high-mass 
star-forming regions \cite{menten} as it lies in close surroundings of
massive protostars. In 2003--07 we used the EVN\footnote{The European
VLBI Network is a joint facility of European, Chinese, South African and
other astronomy institutes funded by their national research councils.}  
to observe 33 methanol sources towards the Galactic plane. We imaged 31
sites and due to the increased sensitivity of EVN we detected a new
morphology that had not been observed so far (\cite{bart09}). 
In at least nine out of 31 sources the methanol maser spots showed 
the {\it ring-like} distribution with major-axes 
from 50 to 400\,mas and the ellipticies between  
0.38 and 0.94. The detailed analysis of maser spots in spectral single channel maps 
revealed that these structures are not consistent with rotating  
Keplerian discs (see Langevelde et al. this issue).

In order to find out more on the nature of the methanol masers in 
high-massive star-forming regions
we have started three projects to investigate other signatures of
high-mass formation such as H\,{\small II} regions, H$_2$O 
maser emission and  MIR objects. The more detailed descriptions of the results 
for the sample of 31 methanol masers can be found 
in \cite{bart09}, \cite{bart10}, while the MIR data are being prepared
for publication elsewhere. 
Here we summarize the results obtained for nine ring-like structures.

\section{Observations and data reduction}
In order to search for radio continuum emission towards the 6.7\,GHz
methanol maser rings 
we used the VLA\footnote{The Very Large Array (VLA) of the National Radio Astronomy Observatory is a
facility of the National Science Foundation operated under cooperative
agreement by Associated Universities, Inc.} at 8.4\,GHz in A configuration 
(AB1250) in the standard continuum mode on 2007 August 18. 
We used phase-referencing in a cycle of 50\,s$+$250\,s for the phase
calibrator and the target, respectively, for a total of 20\,min for each
source. 3C\,286 was used as a flux calibrator and two
phase-calibrators were selected from a standard VLA list: J1851$+$0035 and
J1832$-$1035. The data reduction was carried out following the standard
recipes from AIPS Cookbook Appendix A (NRAO 2007). The final maps were
created using natural weighting and the typical noise level was
$\sim$50\,$\mu$Jy\,beam$^{-1}$ (1$\sigma_{\rm rms}$) and the beam was
0.35\,arcsec$\times$0.25\,arcsec.

We also used VLA to search for the water maser line at 22.23508\,GHz towards the
methanol rings. The observations were carried on 2009 June 4 and
5 (AB1324) in CnB configuration in spectral line mode. We used a single IF
6.25-MHz wide, divided into 128 spectral channels giving a velocity resolution 
of 0.65\,km\,s$^{-1}$. Similarly, as in the previous project, 
3C\,286 was used as the primary flux
density calibrator and J1851$+$0035 and J1832$-$1035 were used as phase
calibrators using the same cycle time, for a total of 29~min for each source.
 The data
reduction was carried out as described in the AIPS Cookbook Appendix B (NRAO 2009). 
The analysis was carried out on naturally weighted images, total size
$77\times77$ arcsec$^2$ and pixel size $0.15\times0.15$ arcsec$^2$. The
typical beam and the noise level per channel were 1.5\,arcsec$\times$0.8\,arcsec
and 4\,mJy\,beam$^{-1}$, respectively. 

We used the Thermal-Region Camera and Spectrograph (T-ReCS) at Gemini
South\footnote{Based on observations obtained at the Gemini Observatory, which is operated
by the Association of Universities for Research in Astronomy, Inc., under a
cooperative agreement with the NSF on behalf of the Gemini partnership: the National Science
Foundation (United States), the Science and Technology Facilities Council (United Kingdom), the
National Research Council (Canada), CONICYT (Chile), the Australian Research
Council (Australia), Ministerio da Ciencia e Tecnologia (Brazil)
and Ministerio de Ciencia, Tecnologia e Innovacion Productiva (Argentina).}
(GS-2009B-Q-7) to image the methanol maser rings at
the two MIR wavelengths (8.7 and 18.3\,$\mu$m). The
observations were taken in several nights in the period of July--August 2009
and the total exposure time were 340\,s at 8.7\,$\mu$m and 360\,s at 
18.3\,$\mu$m. We selected
five objects (G23.389$+$00.185\footnote{The names follow the Galactic
coordinates, G{\it ll.lll$+$bb.bbb}.}, G23.657$-$00.127, G24.634$-$00.324,
G25.411$+$00.105 and G26.596$-$00.024) for MIR observations as they showed 
the brightest emission at the Spitzer IRAC maps among the nine objects. 
 We obtained images with pixel size of 0.089\,arcsec and the 
spatial resolutions of 0.15 and 0.25\,arcsec at 8.7- and 
18.3\,$\mu$m, respectively, after deconvolution. 

\section{Results and discussion}
\begin{figure}
\includegraphics{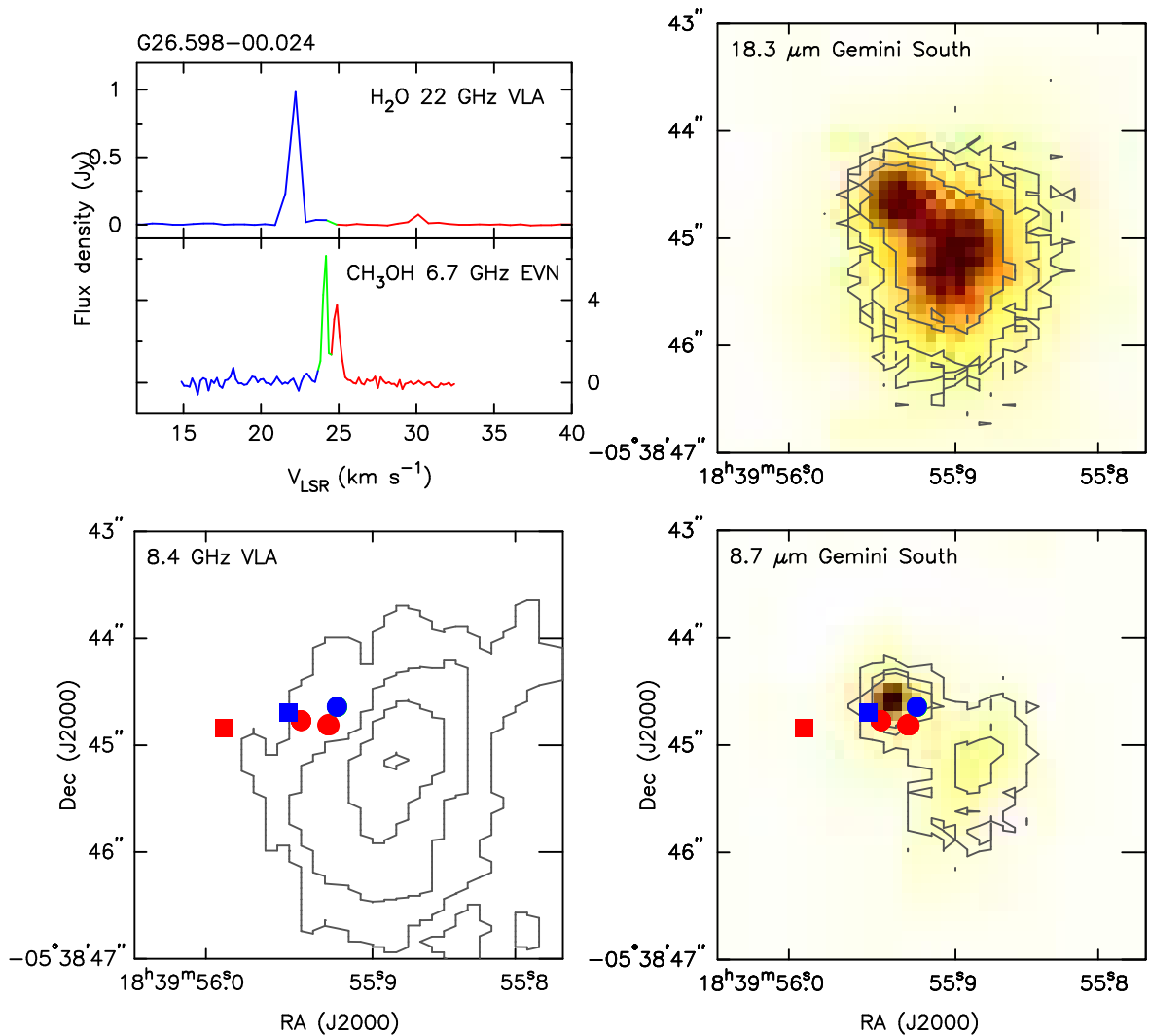}
\caption{An overview of the G26.598$-$00.024 massive star-forming region.
{\bf Top left:} The spectra of 22 GHz water maser and 6.7 GHz methanol maser 
lines taken using VLA and EVN, respectively. {\bf Bottom left:}
Distributions of methanol (circles) and water (squares) maser spots with
colours corresponding to LSR velocities as indicated on the spectra. The
contours trace the emission from the H\,{\small II} region at the levels of
3, 10 and 30 $\times \sigma_{\rm rms}$ detected using VLA at 8.4\,GHz. {\bf Top
and bottom right:} MIR emission towards the target taken using Gemini
South at 8.7 and 18.3\,$\mu$m. The astrometry technique was used at 
8.7\,$\mu$m with a resulting accuracy of 0.18\,arcsec in RA and Dec of the registered
objects.}
\label{fig1}
\end{figure}

We found that 8.4\,GHz continuum emission coincides with the methanol
maser ring of G26.598 $-$00.024. The separation between the
centre of ellipse fitted to the methanol maser spots and the centre of
the H\,{\small II} region is 0.8\,arcsec and the maser is located to the NE
from the central part of the radio continuum object (Fig.~\ref{fig1}). 
Towards the other eight methanol rings we did not
detect any 8.4\,GHz emission stronger than 0.15\,mJy\,beam$^{-1}$
(3$\sigma_{\rm rms}$). That is consistent with previous findings that in general the
6.7\,GHz methanol masers do not have cm-wavelength continuum counterparts.
This implies that
either the maser emission originates prior to the formation of an 
H\,{\small II} region
around a massive proto-star or young star, or alternatively that the young
star is too cool (i.e.~too low-mass) to produce an H\,{\small II} region
(e.g., \cite{beuther}; \cite{pandian}; \cite{phillips}; \cite{walsh}).

Searching for water maser emission resulted in a higher detection rate. 
Five out of nine methanol rings showed the 22\,GHz water line within 0.026\,pc
(the distances are calculated for the near kinematic distances according to
\cite{reid} with systemic velocities taken from \cite{szymczak07}). 
We note that in 80\,per cent of these objects (i.e., 4 out of 5) 
the position angle of the main axis of the
water maser structure is crudely orthogonal ($60^{\rm o}-120^{\rm o}$) 
to the major axis of the methanol ring,  which could be interpreted as an
outflow-disc scenario. However, a careful investigation of 
that hypothesis needs 
multi-epoch observations using VLBI, which we have just started
using the EVN at 22\,GHz. We also note that in four out of five targets 
the water emission is weaker than the methanol emission and that likely 
indicates that the outflows possibly associated with the water maser
emission do not dominate these regions. The lack of water maser 
emission or its weakness might be related to the evolutionary stage
of these regions. According to the model by \cite{beuthershepherd} an 
outflow in massive protostar/young star starts with a collimated jet when a
B-type star forms via accretion through a disk. When an H\,{\small II} region 
forms the wind from a massive young star produces an additional, 
less collimated 
outflow component. Later the H\,{\small II} region expands and a less
collimated wind begins to dominate the whole system.
This is an additional evidence that methanol masers are related to protostars at a
very early stage of evolution -- before the surrounding is ionised and before
the outflow dominates the system. 

Observations using Gemini enabled us to investigate MIR emission
at $\sim$0.15\,arcsec resolution in five selected methanol rings. 
The 8- and 18\,$\mu$m emission appeared as
extended or complex emission in four cases. A
single unresolved MIR source is only found towards G23.657--00.127. We used
an astrometric ''triangulation'' technique on the 8\,$\mu$m data 
to establish that, in four cases, the offsets between the methanol ring
centres and the 8-$\mu$m peaks are within the 1--2$\sigma$ uncertainties
only (0.17--0.21 arcsec) in the 8-$\mu$m positions. This supports the existance
of a relationship between the rings and discs or outflow cavities.

\subsection{G26.598$-$00.024}
We find the case of G26.598$-$00.024 very interesting and we will 
describe it in a little more detail here.  It is the only methanol ring in
out sample which is associated with detectable radio continuum emission. 
 The methanol maser ring is
offset by ca.\,1500\,AU from the central part of the H\,{\small II} region
(for the near kinematic distance of 1.85\,kpc). The peak intensity of the radio
continuum is 4.30\,mJy\,beam$^{-1}$ and its integral flux density 42\,mJy.
Water maser emission appeared in two clusters that lie 1.1 (2035\,AU) 
and 1.6\,arcsec (2960\,AU) from the centre of the H\,{\small II} region 
to the NE, respectively. The distribution of maser spots of both species and
their LSR velocities may suggest a relationship with an outflow originating
from the centre of the  H\,{\small II} region. 
However, the Gemini observations revealed a double structure at 
8 and 18\,$\mu$m with a separation between two peaks of 0.97\,arcsec
(1795\,AU) at a position angle of $+45^{\rm o}$ (as East of North). 
The NE and SW components show flux densities of 0.152 and 0.087\,Jy at
8\,$\mu$m and 2.24 and 0.72\,Jy at 18\,$\mu$m. 
The NE component coincides within 1$\sigma$ with the centre of
the methanol maser ring, while the SW MIR component coincides with the
centre of the H\,{\small II} region. That finding makes it clear that 
the both masers are associated with the NE proto-/young star that has
not ionised its environment yet. Interestingly, that there are no 
masers around the SW object which is very likely more evolved.

\section{Conclusions}
We presented a summary of three projects that studied the methanol maser
emission in the massive star-forming regions. 
Here we focused on the results concerning
the ring-like methanol masers. The maser emission  appears 
in the very early stage
of evolution of a massive proto-star or a young star, when no H\,{\small II}
region exist and outflows are not well formed. It is possible that other
morphologies would appear in later stages of evolution -- 
when outflows form or alternatively, are created by another mechanisms such
as shocks.

\vspace*{1cm}
\noindent
{\bf Acknowledgements}

\noindent
This work was supported by the Polish Ministry of Science and Higher
Education through grant N N203 386937 and the UMK grant 378--A (2010).


\begin{thebibliography}{99}
  \bibitem{bart09}{A.~Bartkiewicz, M.~Szymczak, H.J.~van Langevelde,
A.M.S.~Richards \& Y.M.~Pihlstr\"om 
\emph{The diversity of methanol maser morphologies from VLBI observations},
A\&A, {\bf 2009}, 502, 155}
  \bibitem{bart10}{A.~Bartkiewicz, M.~Szymczak, Y.M.~Pihlstr\"om, H.J.~van Langevelde, 
A.~Brunthaler \& M.J.~Reid  
\emph{VLA observations of water maser towards 6.7\,GHz methanol maser
sources}, A\&A, {\bf 2010}, in press}         
  \bibitem{beuther}{H.~Beuther, A.~Walsh, T.K.~Sridharan, K.M.~Menten \&
F.~Wyrowsky,
\emph{CH$_3$OH and H$_2$O masers in high--mass star--forming regions}, A\&A,
{\bf 2002}, 390, 289--198}
  \bibitem{beuthershepherd} {H.~Beuther \& D.~Shepherd, \emph{Precursors of
UC H\,{\small II} regions and the evolution of massive outflows}, 
arXiv:astro-ph/0502214v1, {\bf 2005}}
  \bibitem{menten}{ K.M.~Menten, \emph{The discovery of new, very strong and
widespread interstellar methanol maser transition}, ApJ, {\bf 1991}, 380,
L75--78}
  \bibitem{pandian}{J.D.~Pandian, E.~Momjian, Y.~Xu, K.M.~Menten \&
P.F.~Goldsmith, \emph{Spectral Energy Distributions of 6.7\,GHz methanol
masers}, A\&A, {\bf 2010}, 522, 8} 
  \bibitem{phillips}{C.J.~Phillips, R.P.~Norris, S.P.~Ellingsen \&
P.M.~McCulloch,
\emph{Methanol masers and their environment at high-resolution}, MNRAS, {\bf
1998}, 300, 1131--1157}
  \bibitem{reid}{M.J.~Reid, K.M.~Menten, X.W.~Zheng, et al.,
\emph{Triginometric parallaxes of massive star-forming regions. VI. Galactic
structure, fundamental parameters, and noncircular motions}, ApJ, {\bf
2009}, 700, 137--148}
  \bibitem{szymczak2002}{M.~Szymczak, A.J.~Kus, G.~Hrynek, A.~Kepa \&
E.~Pazderski,
\emph{6.7\,GHz methanol masers at sites of star formation. A blind survey of
the Galactic plane between $20^{\rm o}\le l \le 40^{\rm o}$ and |b|$\le
0.^{\rm o}$52}, A\&A, {\bf 2002}, 392, 277--286}
  \bibitem{szymczak07}{M.~Szymczak, A.~Bartkiewicz \& A.M.S.~Richards, 
\emph{A multi-transition molecular line study of candidate massive young
stellar objects associated with methanol masers}, A\&A, {\bf 2007}, 468,
617--625}
  \bibitem{walsh}{A.J.~Walsh, M.G.~Burton, A.R.~Hyland \& G.~Robinson,
\emph{Studies of ultracompact H\,{\small II} regions}, MNRAS, {\bf 1998},
301, 640--698}

\end{thebibliography}
\end{document}